# Photoelectrical detection and characterization of divacancy and PL5–PL7 spins in silicon carbide

Naoya Morioka[1,2,*], Tetsuri Nishikawa[1,2], Hiroshi Abe[3], Takeshi Ohshima[3,4], and Norikazu Mizuochi[1,2,5]

1. Institute for Chemical Research, Kyoto University, Uji 611-0011, Japan
2. Center for Spintronics Research Network, Institute for Chemical Research, Kyoto University, Uji 611-0011, Japan
3. National Institutes for Quantum Science and Technology (QST), Takasaki 370-1292, Japan
4. Department of Materials Science, Tohoku University, Sendai 980-8579, Japan
5. International Center for Quantum-field Measurement Systems for Studies of the Universe and Particles (QUP), KEK, Tsukuba, 305-0801, Japan

*e-mail: morioka.naoya.8j@kyoto-u.ac.jp

## Abstract

Photoelectrical detection of magnetic resonance (PDMR) offers a scalable alternative to optical readout of spin defects in semiconductors and is particularly promising for near-infrared (NIR) emitters, where photodetection is often challenging. Here, we demonstrate room-temperature coherent PDMR of PL3 (divacancy), PL5, PL6, and PL7 spins. PL7 and PL5 exhibit notably stronger PDMR than PL6 as opposed to optical detection, indicating higher ionization efficiency and suitability for electrical readout. Rabi oscillation and two-frequency spectroscopy reveal a previously undiscovered secondary resonance of PL7. We determine the zero-field splitting parameters of PL7 and assign the recently reported PL3a defect to PL7. The demonstrated PDMR of these NIR defects constitutes a key advancement toward quantum electronic devices. Also, the clarified spin parameters and ionization characteristics provide a solid foundation for advancing quantum technologies utilizing these defects regardless of the detection schemes.

## Introduction

Electronic spins of point defects in semiconductors have emerged as promising platforms for quantum technologies, including quantum communication [1,2] and sensing [3]. Among them, silicon carbide (SiC) has attracted considerable interest due to the availability of various defects with excellent spin properties even at room temperature. These include the V2 center (silicon vacancy) [4], two of the divacancies labeled PL1 and PL3 [5], and unidentified defects PL5, PL6, and PL7 [5], in the 4H polytype of SiC (4H-SiC). In addition, the maturity of SiC's device fabrication technology and its wafer-scale availability make it a compelling platform for practical room-temperature quantum devices.

Coherent spin-state readout is essential for realizing quantum functionalities based on defects. In addition to the well-established optically detected magnetic resonance (ODMR) technique, electrical readout based on the measurement of spin-dependent photocurrent, known as photoelectrical detection of magnetic resonance (PDMR) [6], has been actively developed. PDMR has demonstrated single-spin sensitivity in both diamond [7] and 4H-SiC [8], with signal contrast comparable to that of ODMR. Notably, PDMR offers a higher signal-to-noise ratio for isolated V2 in 4H-SiC due to more efficient collection of photocurrent compared to photons [8]. The potential for device integration underscores the appeal and advantage of applying PDMR to defects in SiC.

Despite its advantages, the application of PDMR remains limited to only the nitrogen-vacancy center in diamond [6,9,10], V2 in 4H-SiC [8,11,12], and their surrounding nuclear spins [12–14]. Extending this technique to other systems is highly desirable, especially for near-infrared (NIR) defects, where conventional photodetectors often suffer from low efficiency and high dark noise. PL1, PL3, and PL5–7 in 4H-SiC are interesting room-temperature NIR spins. Among them, PL5–7 stand out because of their remarkably higher single-defect ODMR contrasts (10–30%) [15] than those of well-studied V2 (2−6%) [4,8,16], although PL5–7's microscopic structures remain unidentified amid active discussions [17–22], including divacancies in a stacking fault [17], oxygen-vacancy complexes [19,22], and divacancy-carbon antisite complexes [20,22]. These features indicate the potential for achieving high contrast in PDMR as well. Nevertheless, the applicability of PDMR to PL5–7 had remained unclear due to their reported photoionization robustness [23,24] and unidentified charge states.

In this study, we demonstrate room-temperature coherent photoelectrical detection of ensemble PL3 and PL5–7 in 4H-SiC. Notably, PL7 exhibits a stronger PDMR signal than PL6, in contrast to ODMR, indicating that PL7 may serve as a promising system for electronics-based quantum technologies. Encouraged by the promising PDMR response of PL7, we perform an in-depth investigation to identify its spin system, which had remained elusive until now. Using PDMR Rabi-oscillation measurements and two-frequency PDMR measurements, we uncover a previously unreported secondary spin resonance of PL7, overlapping with the resonances of the PL3 divacancy and the recently reported PL3a defects [25]. Our results clarify the identity of PL7 by demonstrating that it is not PL4, as recently proposed [25], but corresponds to PL3a. Based on these findings, we also determine the axial ($D$) and transverse ($E$) zero-field splitting (ZFS) parameters of PL7, which are essential for its application in quantum technologies and for identifying the defect's origin.

## Results and Discussion

### Optical and photoelectrical detection of PL5–7 and PL3

We measured defects in a (0001) high-purity semi-insulating 4H-SiC material. We performed electron irradiation and annealing to increase the density of divacancies to enable comparison of identified divacancies and PL5–7, but PL5–7 may pre-exist, as they reportedly do not generate by electron irradiation [15,18]. PDMR electrodes were fabricated, and microwave (MW) for spin control was applied using a wire approximately parallel to [1$\bar{1}$00]. A schematic of the device is shown in Fig. 1(a). Excitation laser is 905 nm. The photodiode current from the detected fluorescence and the photocurrent in the PDMR device were processed using transimpedance and lock-in amplifiers, with modulation of the MW amplitude by a slow rectangular envelope for signal extraction, as shown in Fig. 1(b) [9,12]. PDMR was measured at 12 V. Further experimental information is provided in Supplemental Material S1 [26].

We present the pulsed ODMR spectra in Fig. 1(c). We scanned the frequency range reported for divacancies (PL1–4) [27] and PL5–7 [5] and observed six strong resonances at 1134.5, 1321.9, 1333.0, 1342.1, 1350.9, and 1374.9 MHz.

The peak at 1134.5 MHz was previously assigned to the lower of two zero-field-split resonance lines of PL3 ($S = 1$), a basal $hk$ divacancy [5,28], where $h$ and $k$ denote the quasi-hexagonal and cubic lattice sites, respectively. Recently, Shafizadeh et al. reported an ODMR peak near and overlapping with PL3, termed PL3a, and interpreted the ODMR signal above 150 K as being dominated by PL3a rather than PL3 [25]. Accordingly, the observed resonance at 1134.5 MHz, denoted PL3LF in this work, is regarded as the lower frequency line of either PL3 or PL3a. We note that, at room temperature, the higher-frequency transitions of PL3 and PL3a have previously been reported to either disappear [5] or remain unobserved [25], respectively. However, as shown below, our measurements indicate that both the previously suggested dominance of PL3a and the assignment of the higher-frequency transition must be revised.

Among the remaining peaks, 1342.1 and 1374.9 MHz are attributed to PL5 ($S = 1$), 1350.9 MHz to PL6 ($S = 1$), and 1333.0 MHz to PL7 [5]. PL6 is an axially symmetric defect ($E = 0$) and therefore exhibits a single resonance peak. In contrast, PL5 is a basal-plane defect with a nonzero $E$, resulting in two resonance lines (denoted PL5LF and PL5HF). PL7 is also believed to be a basal-plane defect with $S = 1$ and a nonzero $E$. However, since only one transition has been observed for PL7 so far, the assignment of $S$ remains tentative, and its $D$ and $E$ parameters are still unknown.

The peak at 1321.9 MHz is assigned to PLX1, an axial *hh* NV¯ ($S = 1$) [29,30]. For PLX1 and PL6, side peaks separated by about 5 MHz are observed, probably attributed to hyperfine coupling with $^{29}$Si nuclear spins [30,31].

In addition to the six major resonances, three minor peaks with opposite signal signs are detected. Two lie between 1260 and 1300 MHz, and an ODMR scan of this region, using an extended integration time, is overlaid and vertically offset in Fig. 1(c), revealing features at 1277.4±0.1 MHz and 1291.8±0.3 MHz, respectively. A third minor feature appears just below PL5HF, as seen in the asymmetric line shape in Fig. 1(c), which is consistent across laser powers.

Over the same frequency range as ODMR, we measured photocurrent response, and we successfully detected pulsed PDMR spectra, as shown in Fig. 1(d). Five strong resonances are observed and assigned to PL3LF (1134.3 MHz), PL5 (1341.8 and 1374.3 MHz), PL6 (1350.6 MHz), and PL7 (1332.4 MHz) based on the correspondence with ODMR. Notably, PLX1 is absent, indicating that PDMR selectively detects PL3 and PL5–7 over NV¯ under 905 nm excitation. Among the three minor peaks observed in ODMR, only the one near 1291 MHz appears in PDMR (1291.3±0.3 MHz). PDMR eliminates overlapping contributions, resulting in cleaner signals and enabling more focused analysis of PL5−7 transitions. This is especially critical around PL7 and PL5 HF, where ODMR is complicated by nearby PLX1 and unidentified resonances, respectively. While the origins of the minor peaks remain unclear, the distinct PDMR peak at 1291 MHz should share similar charge dynamics with the major peaks. Given its resonance frequency [5], it may correspond to PL2, the axial *kk* divacancy, or another unidentified defect.

We now discuss the mechanism of PDMR in PL3 and PL5–7. PDMR relies on spin-dependent photoionization and charge-state cycling, which generate a measurable photocurrent linked to the spin state [6,8,11]. For PL1–4, the charge conversion dynamics have been extensively studied in the context of photostability [23,32–35] and spin-to-charge conversion optical readout [36]. We show here that these dynamics naturally lend themselves to PDMR, as illustrated in Fig. 1(e). Divacancies undergo two-step photoionization from the initial neutral state VV$^0$ to a negatively-charged state VV¯: the first photoabsorption excites the defect from the ground state (GS) to the excited state (ES), and the second promotes a valence-band electron into the defect, resulting in VV¯ and leaving behind a hole. A third photon can then restore the neutral state by ejecting an electron into the conduction band.

Since the photon energy of 905 nm exceeds the zero-phonon line (ZPL) energy (1.12 eV for PL3 [27]), the first ionization threshold (0.8 eV for PL1 [36]), and the charge-state recovery energy (1.321 eV for PL3 [33]), this wavelength is sufficient to drive all transitions required for continuous charge cycling. The spin dependence of the ES lifetime arising from the intersystem crossing renders the initial ionization process spin-selective, resulting in a spin-dependent photocurrent.

In the case of PL5–7, however, their charge states remain unidentified. Moreover, the optical stability of PL5 and PL6 has been attributed to their suppressed ionization [23,24]. Nevertheless, our observation of PDMR signals from PL5–7 indicates that charge-state cycling does occur under 905 nm excitation in a manner analogous to that of divacancies.

In contrast, NV$^-$ behaves differently. Zargaleh et al. reported the disappearance of photoluminescence at excitation energies above ~1.6 eV [37]. Although the mechanism underlying this disappearance is not yet understood, the energy threshold aligns with the theoretical ionization threshold (1.7–1.8 eV) [30]. This suggests that our 905 nm excitation is insufficient to ionize NV$^-$, thereby suppressing the NV$^-$ signal in PDMR. Supporting this, an electron paramagnetic resonance (EPR) study under photoexcitation at 2.1–2.7 eV observed NV$^-$ signals only above 2.3 eV [38], possibly indicating a charge-state recovery. We therefore estimate that a photon energy of approximately 2.3 eV may be required for NV$^-$ detection.

The spectral selectivity in PDMR is useful for an in-depth study of defects, particularly when the emission spectral range is broad (at room temperature) and/or unknown (e.g., PL7 and an unidentified component overlapping at PL5HF), as we show later.

## Ionization efficiency of PL5–7

While the photocurrent in ensemble measurements contains mixed contributions from multiple defect types and background current, resonance-induced relative changes still allow us to infer ionization efficiencies by analyzing the laser power dependence of the spin resonance intensities. Figures 2(a) and (b) show the laser power dependence of the peak intensities in pulsed ODMR and PDMR, respectively, obtained via simultaneous detection. Each signal was normalized to the PL6 intensity at the highest laser power. The ODMR signals saturate at high laser powers, and the intensities follow the order PL6 > PL5 > PL7 ≈ PL3. The order for PL5–7 is consistent with the reported trend in photon count rates and spin contrast for single defects [15]. In contrast, all PDMR signals continue to increase with increasing laser power, and the slope reflects susceptibility to ionization. Notably, PL7 exhibits the strongest PDMR response despite showing weak ODMR, and PL5 and PL3 also surpass PL6 in PDMR, which is the reverse of the trend observed in ODMR.

The ionization probability is generally determined by the ES lifetimes and the ionization cross-section. Since the ES lifetimes of PL3, PL5, and PL6 have been reported to be similar (12–14 ns) [15,39], the higher ionization efficiency of PL3 and PL5 relative to PL6 is likely due to a larger ionization cross-section. The previously reported low photon count rate of PL7 may also be attributed to its higher ionization rate. The origin of the different photoionization cross-sections between PL6 and the other defects remains unresolved; however, possible contributing factors may include differences in ZPL energies, ionization thresholds, and the laser polarization sensitivity between axial and basal defect configurations under $c$-plane excitation.

The enhanced ionization response of PL7 and PL5 suggests that these defects are particularly well suited for PDMR-based readout, offering potential for non-optical quantum technologies. However, for PL7, the lack of known ZFS parameters has so far limited its practical utility. In this study, we address this gap by determining the spin parameters of PL7 through further PDMR-based characterization.

## Spin system analysis by coherent photoelectrical spin detection

To investigate the spin system of PL7, we performed a comparative study of Rabi oscillations of basal defects. Reliable measurements of all relevant spin transitions are essential for accurate spin characterization. In ODMR-based Rabi measurements, the PL5HF signal exhibited instability, likely due to overlapping spectral features (see Fig. 1(c) and [26]). In contrast, PDMR uniquely enabled us to reliably measure Rabi oscillations of PL3 and PL5–7, which was crucial for the spin characterization presented below.

The zero-field resonances of $S = 1$ defects comprise two transitions, $|0\rangle \leftrightarrow |+\rangle$ and $|0\rangle \leftrightarrow |-\rangle$, occurring at frequencies $D + E$ and $D - E$, respectively, where $|\pm\rangle = (|1\rangle \pm |-1\rangle)/\sqrt{2}$. These are driven by the MW field components along the $x$ and $y$ axes of the defect's spin principal frame, respectively. Assuming that the defects lie along the basal Si–C bonds, there are six symmetry-equivalent configurations, as shown in Fig. 3(a). By aligning the MW antenna approximately along $[1\bar{1}00]$ crystal axis, we ensured that the MW field had components only along $[11\bar{2}0]$ and $[0001]$, enabling us to distinguish between the two transitions based on the number and spacing of Rabi frequencies. A detailed derivation is provided in Supplemental Material S3 [26]; briefly, the $|0\rangle \leftrightarrow |+\rangle$ transition yields three equally spaced frequency components, while the $|0\rangle \leftrightarrow |-\rangle$ transition yields two components in a 1:2 frequency ratio.

We examine Rabi oscillations of PL5. PL5HF and PL5LF correspond to the $|0\rangle \leftrightarrow |+\rangle$ and $|0\rangle \leftrightarrow |-\rangle$

transitions of a spin-1 system, respectively. PL5HF [Fig. 3(b)] shows three frequency components, while PL5LF [Fig. 3(c)] contains two. The MW power dependence of the components in PL5HF [Fig. 3(f)] confirms that three frequencies are almost equally spaced, and that of PL5LF [Fig. 3(g)] verifies an almost 1:2 frequency ratio. These observations agree with theoretical expectations, supporting the assignments and serving as a reference for subsequent analysis.

PL7, shown in Fig. 3(d), exhibits a Rabi oscillation with three frequency components and comparable power dependence to PL5HF [Fig. 3(f)]. This strongly suggests that PL7 corresponds to the $|0\rangle \leftrightarrow |+\rangle$ transition of another $S = 1$ defect.

In contrast, PL3LF shows an oscillation that cannot be described by two or three components [Fig. 3(e)]. A fitting analysis reveals four distinct frequency components. As shown in Fig. 3(g), two of them match those observed for PL5LF. The remaining two are also close in value. This indicates that PL3LF consists of overlapping $|0\rangle \leftrightarrow |-\rangle$ transitions from two distinct $S = 1$ defects, PL3 and PL3a, coexisting at room temperature, in contrast with previous interpretations that PL3a dominates above 150 K [25].

The observation of both PL7 and PL3LF in PDMR suggests a shared ionization mechanism. Since PL7 shows only the $|0\rangle \leftrightarrow |+\rangle$ transition and PL3LF two $|0\rangle \leftrightarrow |-\rangle$ transitions, one may originate from PL7. Given that PL3's higher-frequency transition at 20 K (1304 MHz [5]), which decreases with increasing temperature [25], is lower than PL7's room-temperature resonance, we exclude PL3 and hypothesize that PL3a and PL7 form a pair.

To test this hypothesis, we performed two-frequency PDMR measurements, adapting a technique that previously demonstrated distinction of orientations of nitrogen-vacancy centers in diamond via ODMR [40]. This method uses a pair of MW pulses (MW1 and MW2) at frequencies $f_1$ and $f_2$, respectively (Fig. 4(a)). MW1 drives one of the $|0\rangle \leftrightarrow |\pm\rangle$ transitions of a defect. If MW2 addresses the corresponding $|0\rangle \leftrightarrow |\mp\rangle$ transition of the same defect, MW1 alters the signal detected via MW2. Transitions from unrelated defects remain unaffected. Lock-in detection with on-off modulation of MW1 selectively isolates signal components originating from the spin addressed by MW1 in the sweep of MW2.

As an experimental demonstration [Fig. 4(b)], we set $f_1$ to PL5HF, and found a response only at PL5HF, confirming selective detection of PL5 in the spin ensemble. For the correspondence of PL3a and PL7, setting $f_1$ to PL3LF revealed a response only at PL7. Reversing roles ($f_1$ at PL7) produced a corresponding signal at PL3LF. These results confirm that PL3a spin resonance is assigned to PL7.

From this pairing, we extract the room-temperature ZFS parameters of PL7 as $D = 1233.6$ MHz and $E = 99.0$ MHz. Although a recent study proposed reassigning PL7 to PL4 [25], our findings strongly support that PL7 and PL3a are the same defect.

## Conclusions

In summary, we demonstrated room-temperature coherent and selective PDMR of PL3, PL5, PL6, and PL7 over NV$^-$ and unidentified transitions. Among these, PL7 and PL5 exhibit significantly stronger PDMR signals than PL6, suggesting higher ionization efficiency and their suitability for PDMR readout. We identified PL7 as an $S = 1$ spin defect and determined its ZFS parameters using PDMR, leading to the assignment of PL3a to PL7.

The clarified spin characteristics of PL7 provide a solid foundation for its practical utilization. Moreover, the obtained ZFS values and ionization behavior provide essential benchmarks for theoretical modeling of PL5–7, whose microscopic structures remain unidentified and challenging for first-principles calculations, and could ultimately enable defect engineering. These findings are broadly valuable for the development of quantum technologies based on PL5–7 defects, regardless of the readout scheme.

Among possible approaches, electrical spin detection offers particular advantages for integration and miniaturization. While PL5–7 are already known for their excellent optical properties, our results based on the developed electrical readout technique reveal that they also possess promising characteristics for electrical detection. Applying sensitive single-spin PDMR to PL5–7 would significantly improve the efficiency of electrical spin readout. Taken together, these insights lay the groundwork for realizing scalable quantum electronics technologies based on defect spins in silicon carbide.

## Acknowledgments

We thank Hiroki Morishita, Masanori Fujiwara, Nguyen Tien Son, Ivan G. Ivanov, and Danial Shafizade for the fruitful discussions. We also thank Teruo Ono, Takahiro Moriyama, and Yoichi Shiota for their support in the device fabrication. This work was partly supported by JSPS KAKENHI (Grant Nos. JP25K01262, JP23K22796, and JP21K20502), JST PRESTO Grant Number JPMJPR245C, MEXT Q-LEAP Grant Number JPMXS0118067395, research grant from The Mazda Foundation, research grant from Murata Science and Education Foundation, Asahi Glass Foundation, JST SPRING (Grant Number JPMJSP2110), the Spintronics Research Network of Japan, "Advanced

Research Infrastructure for Materials and Nanotechnology in Japan (ARIM)" of the Ministry of Education, Culture, Sports, Science and Technology (MEXT) (Proposal Number JPMXP12 24KT2474), and JST ASPIRE Grant Number JPMJAP24C1.

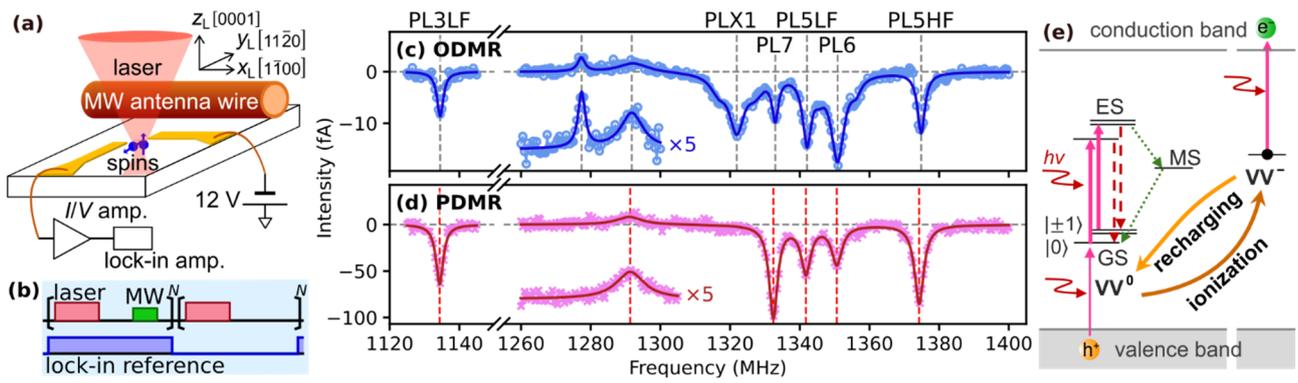

Figure 1. (a) Device structure and electrical configuration for PDMR. (b) ODMR and PDMR pulse sequences. (c) ODMR spectra measured at 91 mW laser. A scan of 1260–1300 MHz with extended integration time at 27 mW is overlaid. (d) PDMR spectra at 91 mW. An extended-integration scan (1260–1310 MHz) at 91 mW is overlaid. Baseline was subtracted in (c) and (d). (e) PDMR process in divacancies. Straight arrows indicate optical excitation (solid), radiative decay (dashed), and major spin-dependent intersystem crossing pathways (dotted).

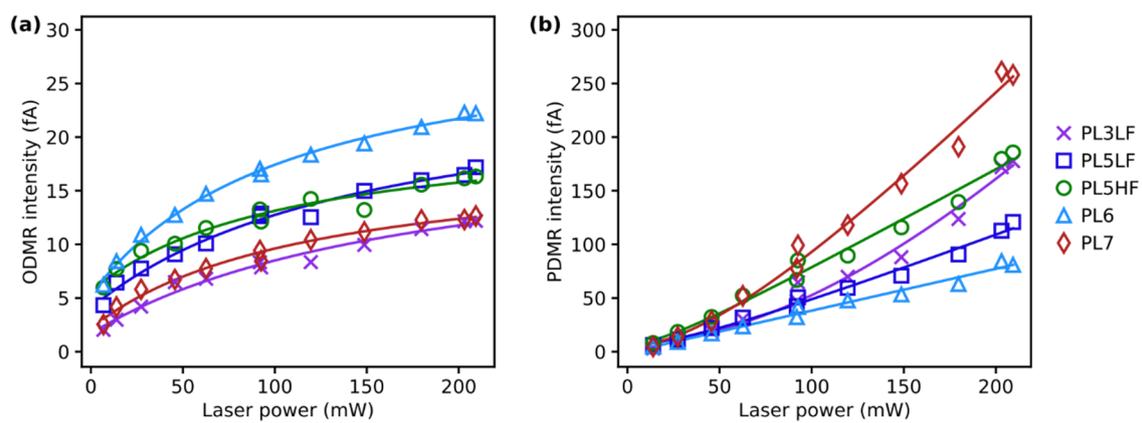

Figure 2. Laser power dependence of (a) ODMR and (b) PDMR peak intensities. Lines are guides to the eye.

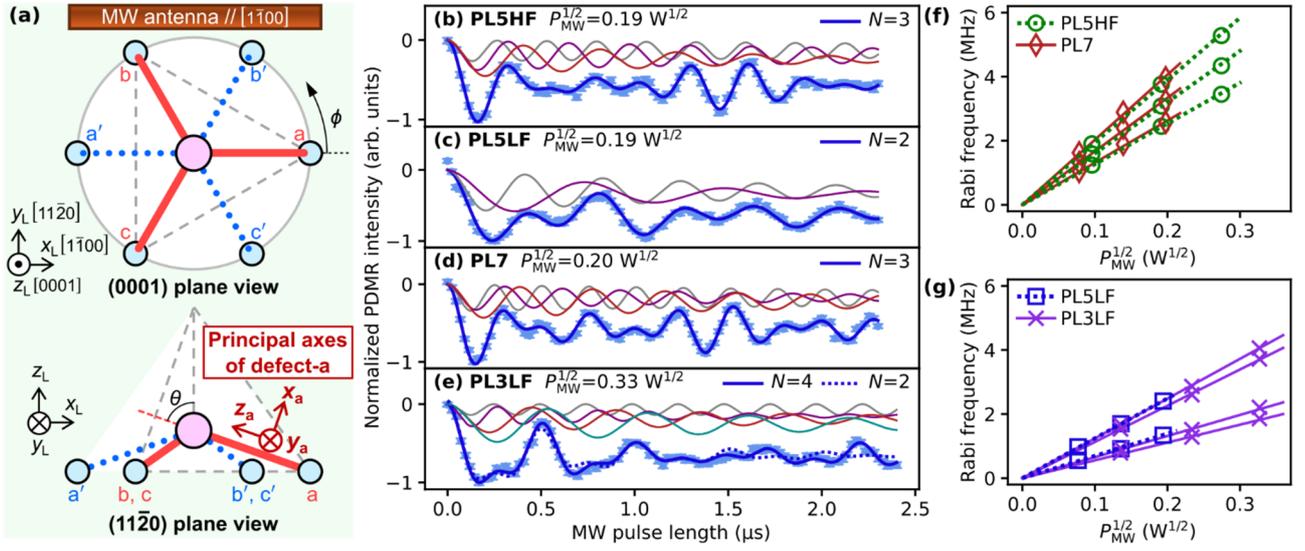

Figure 3. (a) Six configurations (a, b, c, a′, b′, c′) of basal defects. Principal axes are shown for defect-a as an example. (b–e) PDMR-detected Rabi oscillations of (b) PL5HF, (c) PL5LF, (d) PL7, and (e) PL3LF. Thick lines represent fits assuming $N$ frequency components, and thin lines show each component ($N$=4 for PL3LF). Linear background from microwave noise [12] has been subtracted. (f, g) MW power dependence of the Rabi frequencies for: (f) PL5HF and PL7, and (g) PL5LF and PL3LF. Lines are fits to square-root power dependence.

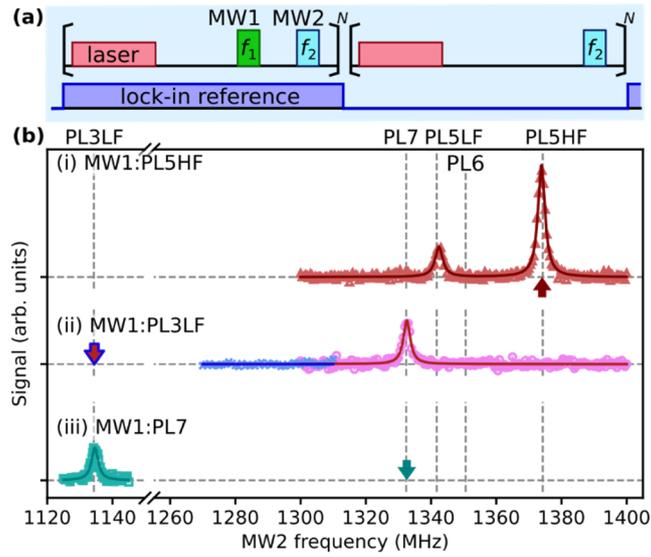

Figure 4. (a) Pulse sequence of two-frequency PDMR. MW1 is fixed, MW2 is swept. (b) Spectra with MW1 at (i) PL5HF, (ii) PL3LF, and (iii) PL7. Thick arrows indicate MW1 frequencies. Baseline signal arising from MW1 is subtracted.